\documentclass[preprintnumbers,amsmath,11pt,amssymb,floatfix,superscriptaddress,nofootinbib]{article}

\def\mytitle#1{\setcounter{equation}{0}
\setcounter{footnote}{0}
\begin{flushleft}\Large\textbf{#1}\end{flushleft}
\vspace{0.20cm}}
\def\myname#1{\leftline{{\large #1}}\vspace{-0.13cm}}
\def\myplace#1#2{\small\begin{flushleft}\textit{#1}\\
\texttt{#2}\end{flushleft}}

\def\myclassification#1{\small\noindent
Keywords : Dark Energy, Scale Factor, Redshift parametrization.\\

#1\vspace{0.5cm}}
\usepackage{graphicx}
\usepackage{amsmath}
\begin{document}
\mytitle{Evolution of Universe as a Homogeneous System: Changes of Scale Factors with Different Dark Energy Equation of States}

\myname{$Promila ~Biswas^{*}$\footnote{promilabiswas8@gmail.com} and  $Ritabrata~
Biswas^{\dag}$\footnote{biswas.ritabrata@gmail.com}}
\myplace{Department of Mathematics, The University of Burdwan, Golapbag Academic Complex,
City : Burdwan-713104, Dist. : Purba Barddhaman, State : West Bengal, India.} {}
 
\begin{abstract}
We analyze the universe as a thermodynamic system, homogeneously filled up by exotic matters popularly named as dark energies. Different dark energy models are chosen. We start with the equation of continuity and derive the time and scale factor relations for different EoSs of different dark energy models. To do the time-scale factor relation analysis, nature of dependences on different dark energy modeling parameters have been studied. For this, the help of different plots are used. In general, different dark energies show different properties while occurrences of future singularities are considered. Those properties can be supported by the graphical analysis of their cosmic time-scale factor studies.
\end{abstract}

\myclassification{PACS No.: 98.80.-k, 95.35.+d, 95.36.+X, 98.80.Jk}

\section{Introduction}
Einstein proposed his famous field equation $G_{\mu \nu} = 8 \pi G T_{\mu \nu}$ to relate the curvature of space-time and the stress energy originated from the presence of the matter present in the said space-time. Those days, the simplest and the most elegant assumption was to think our universe as a homogeneous and isotropic system on a large scale. Firstly, the solution of Einstein field equation had predicted our universe to be an expanding one but to support and establish Einstein's personal belief that our universe is static, Einstein has added a cosmological constant ($\Lambda$) term to the geometric part of his equation such that he has forcefully made his model of universe to be a static one. Cosmological constant, if is thought to be added to left hand side of Einstein's field of equation, can be treated as the first time ever modification of gravity. On the other hand, if it is taken to be an additional term on the stress energy part, it can be treated as the inclusion of exotic matter for the very first time of cosmological studies.

However, soon after Hubble's discovery of the fact that distant galaxies are red shifted, i.e., our universe is expanding, Einstein aborted the cosmological constant term. But recent day obsevations of late time cosmic acceleration \cite{cosmic_acceleration_paper} is best supported by the models where cosmological constant($=-1$) is treated to be responsible for such a late time accelerated expansion \cite{Padmanabhan_Cosmological_Constant_best}.

It is popular to take the Friedmann-Lemaitre-Robertson-Walker (FLRW) metric to describe the homogeneous and isotropic universe. The Einstein's equation derived for the metric shows the violation of strong energy condition, i.e., $\rho+3p < 0$ to have late time acclerated expansion, where $\rho$ and $p$ are energy density and pressure respectively. Violation of weak energy condition happens if $\rho+p<0$. For all these scenario we will take the EoS for an exotic fluid homogeneously distributed all over in our universe and exerting negative pressure as $p= \omega\rho$. This equation signifies radiation($\omega= \frac{1}{3}$), dust($\omega=0$), quintenssence($-1< \omega < -\frac{1}{3}$) and phantom($\omega<-1$) as we change the EoS parameter accordingly. All $p, \omega$ and $\rho$ can be treated as functions of redshift $z$. In such a case, different red shift parametrizations of EoS parameter have been proposed.

Two mainstream families of red shift parametrizations are there, viz.,
\begin{itemize}
\item[(i)] Family I : $ \omega(z)= \omega_0+ \omega_1  (\frac{z}{1+z})^n $ and
\item[(ii)] Family II : $ \omega(z)= \omega_0+ \omega_1  \frac{z}{(1+z)^n} $
\end{itemize}
where, $\omega_0$ and $\omega_1$ are two undecided parameters, $n$ is a natural number. Some particular $`n$' cases for both family I and II are discussed below along with some dark energy model having non linear dependence of $p$ on $\rho$.\\

{\scshape{\LARGE Model I :}} \textbf{Linear  parametrization :}

If someone follows the history of the studies of dark energy models, the simplest ever model of redshift parametrizaton was the linear parametrization. The EoS is derived for $n=0$ of family II, i.e., $\omega(z)=\omega_0+\omega_1z$ \cite{Linear_Parametrization}. In this work, using Existing data (Mainly 192 SNeIa) authors have shown that the extreme forms of dynamical behaviour can be ruled out and at the $68\%$ confidence level. During the study, models for which $\omega(z)$ starts from $-0.4$ and above at $z>1$ or the models which are asymptotic to values of $\omega(z)>-0.85$ at $z<0.2$ have been considered. Here $\omega =− \frac{1}{3} $ and $\omega_1=−0.9$ with $z<1$ when Einstein gravity has been considered. However, This grows increasingly unsuitable for $z\gg 1.$. Authors argued against the previous claims in the literature of evolving dark energy, such as ``dark energy metamorphosis". Upadhye-Ishak-Steinhardt \cite{Upadhye_Ishak_Steinhardt} parametrization can avoid this problem  uncertainities in measuring $\omega$ and $\frac{d\omega}{dz}$ at $z=0$ has been resolved in \cite{Upadhye_Ishak_Steinhardt}. So we have the energy density pressure relation for Linear Parametrization as 
\begin{equation}\label{linear_redshift_parametrization}
p(z)=\{\omega_0+\omega_1z\}\rho(z)
\end{equation}\\

{\scshape{\LARGE Model II :}} \textbf{CPL parametrization : }

For $n=1$, both  the  families I and II lead to the same parametrization, $\omega(z)= \omega_0+ \omega_1  (\frac{z}{1+z})$. This ansatz was first discussed by Chevallier and  Polarski \cite{Chevallier_and_Polarski} and later studied more elaborately by Linder \cite{Linder_E_V} . In  Einstein gravity, the best fit values for this model while fitting with the SNIa gold data set are $\omega_0=−1.58$  and $\omega_1 = 3.29.$ This  parametrization will be shortly named as ``CPL Parametrization" after the proposer Chevallier-Polarski-Lindler. There are literatures which support that CPL parametrization has the quantity to catch the dynamics of many DE models and in particular the dynamics of the step like ones \cite{Linden_S_Virey_J}. Here the pressure energy density relation turns to be
\begin{equation}\label{CPL_redshift_parametrization}
p(z)=\left\{\omega_0+\frac{\omega_1 z}{1+z}\right\}\rho(z)
\end{equation}

{\scshape{\LARGE Model III :}} \textbf{JBP parametrization:}

For family II, $n = 2$ gives the parametrization $\omega(z) = \omega_0 +\frac{\omega_1 z}{(1+z)^2}$ . A fairly rapid evolution of this EoS is allowed so that $\omega(z) \geq -\frac{1}{2} $ at $z > 0.5$ is consistent with the supernovae observation in Einstein gravity. We will call this parametrization as JBP (Jassal-Bagla-Padmanabhan) \cite{Jassal_Bagla_Padmanabhan} parametrization. In this corresponding work, authors have considered constraint distances and the main constraint from CMB observations. No perturbation in dark energy is considered. Concerned analysis of models and observations shows that the allowed variation of $\rho^{DE}(z)$ is strongly constrained by a combination of WMAP data and observations of high redshift supernova. So we have for JBP Parametrization,
\begin{equation}\label{JBP_redshift_parametrization}
p(z)=\left\{\omega_0+\frac{\omega_1z}{(1+z)^2}\right\}\rho(z)
\end{equation}

{\scshape{\LARGE Model IV :}} \textbf{Log parametrization:} 

The EoS in this case is given as, $\omega(z) = \left\{\omega_0+\omega_1\ln(1+z)\right\}$
 \cite{Efstathiou_1999}.  This evolution form of EoS is valid for $z < 4$. $\omega_1$ is a small number which can be determined by the observations. The minimum value of $\omega_1$ is approximately $−0.14$ and $\omega_0 \geq −1.$ We will call it log parametrization. 
\begin{equation}\label{Logarithmic_redshift_parametrization}
p(z)=\left\{\omega_0+\omega_1\ln(1+z)\right\}\rho(z)
\end{equation}

The combination of supernova and CMB anisotropy measurements can break the degeneracy between $\Omega_{\lambda}$ and $\Omega_{m}$ if the initial functions are assumed to be adiabatic and characterised by a smooth fluctuation spectrum. This method applied to recent supernova and CMB data suggests a nearby spatially flat universe dominated by a cosmological term with $\Omega_{\lambda} \approx 0.65$. Keeping these drawbacks in mind Log parametrization is proposed.

{\scshape{\LARGE Model V :}} \textbf{ASSS(Alam − Sahni − Saini − Starobinski) parametrization:}

For this redshift parametrization, the EoS is given as $\omega(z) = \left\{-1+\frac{(1+z)}{3}\frac{A_1+2A_2(1+z)}{A_0+2A_1(1+z)+A_2(1+z)^2}\right\}$
\cite{Alam_2004a, Alam_2004b}. This ansatz is exactly the cosmological constant $\Lambda = −1$ for $A_1 = A_2 = 0$ and DE models with $\omega = -\frac{2}{3}$ for $A_0 = A_2 = 0$ and $\omega = − \frac{1}{3}$ for $A_0 = A_1 = 0.$ It has also been found to give excellent results for DE models in which the equation of state varies with time including quintessence, Chaplygin gas, etc. The best fit values of $A_1$ and $A_2$ are $A_1 = −4.16$ and $A_2 = 1.67$ for the SNIa Gold data set. We will call this parametrization as ASSS(Alam − Sahni − Saini − Starobinski) parametrization.
\begin{equation}\label{ASSS_redshift_parametrization}
p(z)=\left\{-1+\frac{(1+z)}{3}\frac{A_1+2A_2(1+z)}{A_0+2A_1(1+z)+A_2(1+z)^2}\right\}\rho(z)
\end{equation}
Besides these linear EoS, there exist some non-linear equations of pressure and density like Chaplygin gas family (Chaplygin gas, modified Chaplygin gas, generalised cosmic Chaplygin gas etc are candidates of his family)\\
{\scshape{\LARGE Model VI :}} \textbf{Generalized Cosmic Chaplygin-Gas (GCCG) Model :}

We introduce here some generalized from the cosmic Chaplygin-gas model \cite{Pedro F_González_Díaz} that also contains an adjustable initial parameter $\omega$.  In particular , we shall consider a generalized gas whose equation of state reduces to that of current Chaplygin unified models for dark matter and energy in the limit $\omega \longrightarrow 0$ and satisfies the following conditions : 
\begin{itemize}
\item[i] It becomes a de Sitter fluid at late time and when $\omega=-1$ .
\item[ii] It reduces to $p = \omega \rho$ in the limit that the Chaplygin parameter $A \rightarrow 0$ .
\item[iii] It also reduces to the equation of state of current Chaplygin unified dark matter models at high energy density,and 
\item[iv] the evoluion of density perturbations derived from the chosen equation of state becomes free from the above mentioned pathalogical behaviour of the matter power spectrum for physically reasonable values of the involved parameters, at late time. We shall see that these generalizations retain a big rip if they also show unphysical oscillations and exponential enlargement leading to instability [i.e if they do not satisfy condition (iv)]
\end{itemize}
\begin{equation}\label{GCCG_EoS}
p = -\rho^{-\alpha}\left[C+\left\{\rho^{1+\alpha}-C\right\}^{-\omega}\right] 
\end{equation} 

We wish to check the nature of dependence of $t$ and $a(t)$ as functions of different dark energy model parameters. This may leasd us to a basic idea regarding the singularities to arise in different models.

The studies of scale factor and time relation have been studied since a long time. The authors of \cite{Evolution_of_the_Scale_Factor_with_a_Variable} have studied it almost for the first time. They have considered the presence of a variable cosmological term $\Lambda$ and extended existence of treatments by adopting a fairly general equation of state for ordinary matter. For $\Lambda \propto t^{-l}$, they have obtained some exact solution of $a(t)$. Thermodynamic constraints on a varying cosmological constant like term from holographic equipartition law with a paper law corrected entropy was studied \cite{Thermodynamic_constraints}.

If we follow the chronology of the universe then we predict a radiation dominated era which might have occured after inflation (about 47,000years after the big bang) and follows the equation $a(t)\propto t^\frac{1}{2}$\cite{Alam_2004a}. Between about $47,000$~years and $9.8$ billion years after the big bang, the energy density of matter exceeded both energy density of rotation and exotic matter. For a matter derivated universe the evolution of scale factorcan be calculated as $a(t)\propto t^\frac{2}{3}$\cite{Thompson_learning, Ryden_Introduction_to_Cosmology}.

The dark energy dominated era \cite{Riess_et} began after the matter dominated era, i.e., when the universe was about $9.8$ billion years old. For dark energy dominated universe, the evolution of the scale factor is generally formed by solving the relaton $a(t) \propto exp(Ht)$. 

This exponential dependence on time makes the geometry of the present universe's space-time identical to the de Sitter universe, and only holds for a positive sign of the cosmological constant, the sign that was observed to be realised in nature is of the order of $9.44 \times 10^{-27}$ kg $m^{-3}$ and the age of the universe is of the order of $13.79$ billion years. We wish to study the scale factor- time relations for different dark energy models.

We can find several literature where authors have tried to explore more general conditions under which a dark energy density that increases with time can avoid a future singularity \cite{The_Little_Rip}.

In \cite{Borrow_Class_Quantum_Gravity}, authors constructed an explicit example by seeking the time interval $0 < t < t_s$, a solution for the scale factor $a(t)$ of form
\begin{align*}
a(t)=1 + Bt^q + c(t_s - t)^n
\end{align*}
where $B > 0$, $q > 0$, $c$ and $n > 0$ are free constants to be determined. Fixing the zero of time by requiring $a(0)=0$, $Ct_s^n= -1$,
the scale factor is obtained as,
\begin{align*}
a(t)= \left(\frac{t}{t_s}\right)^q (a_s -1) + 1-\left(1 -\frac{t}{t_s}\right)^n
\end{align*}
where $a(s) \equiv a(t_s)$. Hence, as $t \rightarrow t_s$ from below, 
we have,
\begin{align*}
\ddot{a} \rightarrow q(q-1)Bt^{q-2}-\frac{n(n-1)}{t_s^2\left(1-\frac{t}{t_s}\right)^{2-n}} \longrightarrow -\infty
\end{align*}
whenever $1 < n < 2$ and $0 < q \leq 1$;
a solution with similar properties which expands from a de Sitter past state also exists, with
\begin{align*}
a(t) = a_s -1 +exp\{\lambda(t_s -t)\}-\left(1-\frac{t}{t_s}\right)^n
\end{align*} 
with $\lambda > 0$ constant and $1 < n < 2$.

Following \cite{More_General_Sudden_Singularities} and \cite{Goriely_A_Hyde_C}, we can obtain a generalisation of the solution as
\begin{align*}
a(t) = \left(\frac{t}{t_s}\right)^q (a_s -1) + 1 - (t_s - t)^n\left[\sum_{j=0}^{\infty} \sum_{k=0}^{N_j}a_{jk}(t_s - t)^{\frac{j}{Q}} (Log^k[t_s - t])\right]
\end{align*}
and as $t \rightarrow t_s$ if we choose $1 < n < 2$, $0 < q < 1$, we have $a \rightarrow a_s$.

In this paper we are motivated find the relation of $a(t) - t$. If we will not be able to give a concrete analytical solution. We will provide a numerical solution via graphs. 
In the next chapter we calculate the relations between time $t$ and the scale factor $a(t)$. We plot them graphically and analyse. At the last we give a brief discussion.

\section{Basic Calculations and Graphical Analysis}
The Friedmann-Lemaitre-Robertson-Walker metric for a homogeneous isotropic universe is provided by in a different form : 
\begin{equation}\label{FLRW_Metric}
ds^2 = \frac{a^2(t)}{\left(1+\frac{k_0r^2}{4}\right)^2}\left[dr^2+r^2d\theta^2 + r^2 sin^2 \theta d\phi^2 \right] -dt^2
\end{equation}
Here $k_0=0,1,-1,$ indicating the spatial curvature constants. Now, the field equations are given by \cite{Raychaudhuri_Banerjee_Banerjee} 
\begin{equation}\label{Raychaudhuri_equation}
\frac{2\ddot{a}}{a} + \frac{\dot{a}^2 + k_0}{a^2} = -kp + \Lambda \\ ~~~and~~~3\left(\frac{\dot{a}^2 + k_0 }{a^2}\right) = k\rho + \Lambda.
\end{equation}
The energy conservation equation can be written as
\begin{equation}\label{energy_conservation_equation}
\dot{\rho} + 3\frac{\dot{a}}{a}(\rho + p) = 0
\end{equation}
We digress here slightly to deduce some consequences of the field equations (\ref{FLRW_Metric}) and (\ref{Raychaudhuri_equation}). We get clearly from these equation
\begin{equation}\label{field_equation_I}
\ddot{a} = -\frac{k}{6}\left( \rho + p \right)a + \frac{\Lambda}{3}a.
\end{equation}
The equation (\ref{field_equation_I}) is sometimes referred as the Raychaudhuri equation \cite{Raychaudhuri_equation_ref1,Raychaudhuri_equation_ref2}. It provides the cosmic acceleration which is governed by forces on the right hand side of (\ref{field_equation_I}).

The field equations yields
\begin{equation}\label{field_equation_II}
\frac{1}{2}\dot{a^2}- \left(\frac{4\pi}{3}\rho + \frac{\Lambda}{6}\right) a^2 = -\frac{k_0}{2}
\end{equation}
The equation (\ref{field_equation_II}) stands for the conservation equation of the total energy of the universe with kinetic energy $ \frac{1}{2}\dot{a}^2 $ and the gravitational potential energy as the second part on the left hand side of (\ref{field_equation_II}).

from (\ref{FLRW_Metric}), (\ref{Raychaudhuri_equation}), (\ref{energy_conservation_equation}) and (\ref{field_equation_II}) to obtain the general solution of the system. We define a function 
\begin{equation}
M(\rho) = exp \left[ \int \frac{d\rho}{\rho + p} \right] > 0
\end{equation}
Considering the pressure $p$ as a function of density, i.e, $p (\rho)$. We get
\begin{equation}
\frac{dM(\rho)}{d\rho} = \frac{M(\rho)}{\rho + p} > 0
\end{equation}
The conservation equation (\ref{energy_conservation_equation}) reduces to
\begin{equation}\label{Mass_Scale_Factor_Relation}
\frac{d}{dt}\left[ln M(\rho) + ln a^3 \right] = 0 
\Rightarrow M(\rho)a^3 =m_0
\end{equation}
Now from \textbf{Linear Parametrization} equation (\ref{linear_redshift_parametrization}) and equation (\ref{Mass_Scale_Factor_Relation}) we get,
\begin{equation}\label{linear_II}
M = \rho^ \frac{1}{1+ \omega_0 + \omega_1 z} = m_0a^{-3}  \Rightarrow \rho = (m_0a^{-3})^{(1+ \omega_0 + \omega_1 z)} \Rightarrow p(a)= \omega(a)(m_0a^{-3})^{(1+ \omega_0 + \omega_1 z)} 
\end{equation}
Now from equation (\ref{field_equation_II}) and (\ref{linear_II})

\begin{equation}\label{adot_for_LP}
\dot{a}^2 = 2\left\{\frac{4\pi}{3} (m_0a^{-3})^{(1+ \omega_0 + \omega_1 z)}  + \frac{\Lambda}{6}\right\} a^2 - k_0 \Rightarrow \left(\frac{da}{dt}\right) = \left[ 2\left\{\frac{4\pi}{3} (m_0a^{-3})^{(1+ \omega_0 + \omega_1 z)}  + \frac{\Lambda}{6}\right\} a^2 - k_0 \right]^{ \frac{1}{2}} ~~.
\end{equation}
So, solving $t$ for $a(t)$ we get, 
\begin{equation}\label{t-a_for_LP}
 \Rightarrow t - t_0 = \int \left[ 2\left\{\frac{4\pi}{3} (m_0a^{-3})^{(1+ \omega_0 + \omega_1 z)}  + \frac{\Lambda}{6}\right\} a^2 - k_0 \right]^{-\frac{1}{2}} da
\end{equation}
To find analytic solution of (\ref{t-a_for_LP}) we assume $\omega_0= -1, \omega_1= 0, m_0= 1$ we have for $k_0= 0$
\begin{equation}
t(a)= \sqrt{\frac{3}{8\pi - 1}}~~lna
\end{equation}
for $k_0= 1$
\begin{equation}
t(a)= \sqrt{\frac{3}{8\pi - 1}}~~ln~~\frac{a \sqrt{8\pi -1 } + \sqrt{a^2(8\pi - 1) -3}}{a\sqrt{8\pi - 1} +  \sqrt{8\pi - 4}} 
\end{equation}
for $k_0= -1$
\begin{equation}
t(a)= \sqrt{\frac{3}{8\pi - 1}}~~\left( \sinh^{-1}\left[\sqrt{\frac{8\pi - 1}{3}}\right] + \sinh^{-1}\left[a \sqrt{\frac{8\pi - 1}{3}}\right] \right)
\end{equation}
Now we will solve the equation (\ref{t-a_for_LP}) numerically and plot, We plot $t$ vs $a(t)$ for $k_0 = 0$  and $1 , -1$ cases in Figs. 1(a)-1(c), respectively .\\
\begin{figure}[h!]
\begin{center}
(a)~~~~~~~~~~~~~~~~~~~~~~~~~~~~~~~~~~~~~~~~~~~~(b)~~~~~~~~~~~~~~~~~~~~~~~~~~~~~~~~~~~~~~~~~~~(c)
\includegraphics[height=1.7in, width=2.2in]{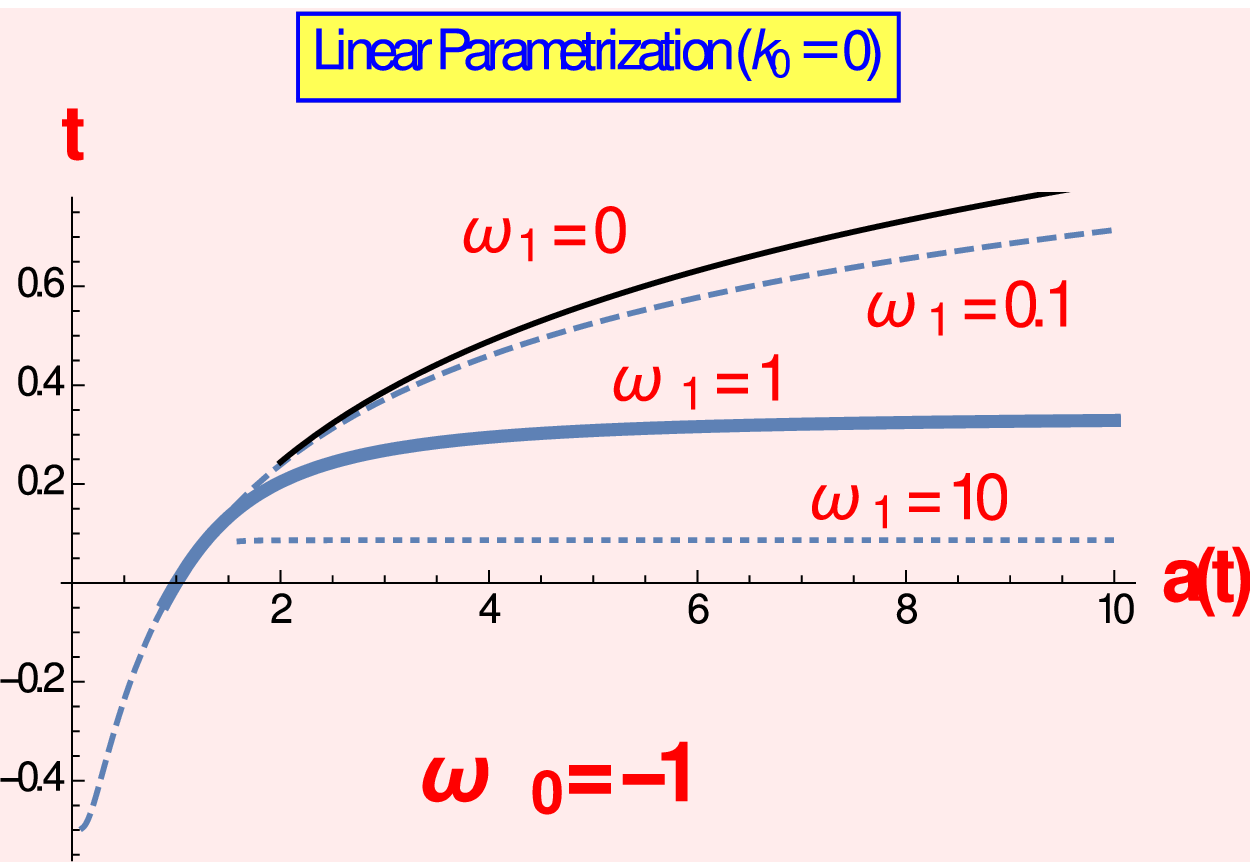}~~~~~\includegraphics[height=1.7in, width=2.2in]{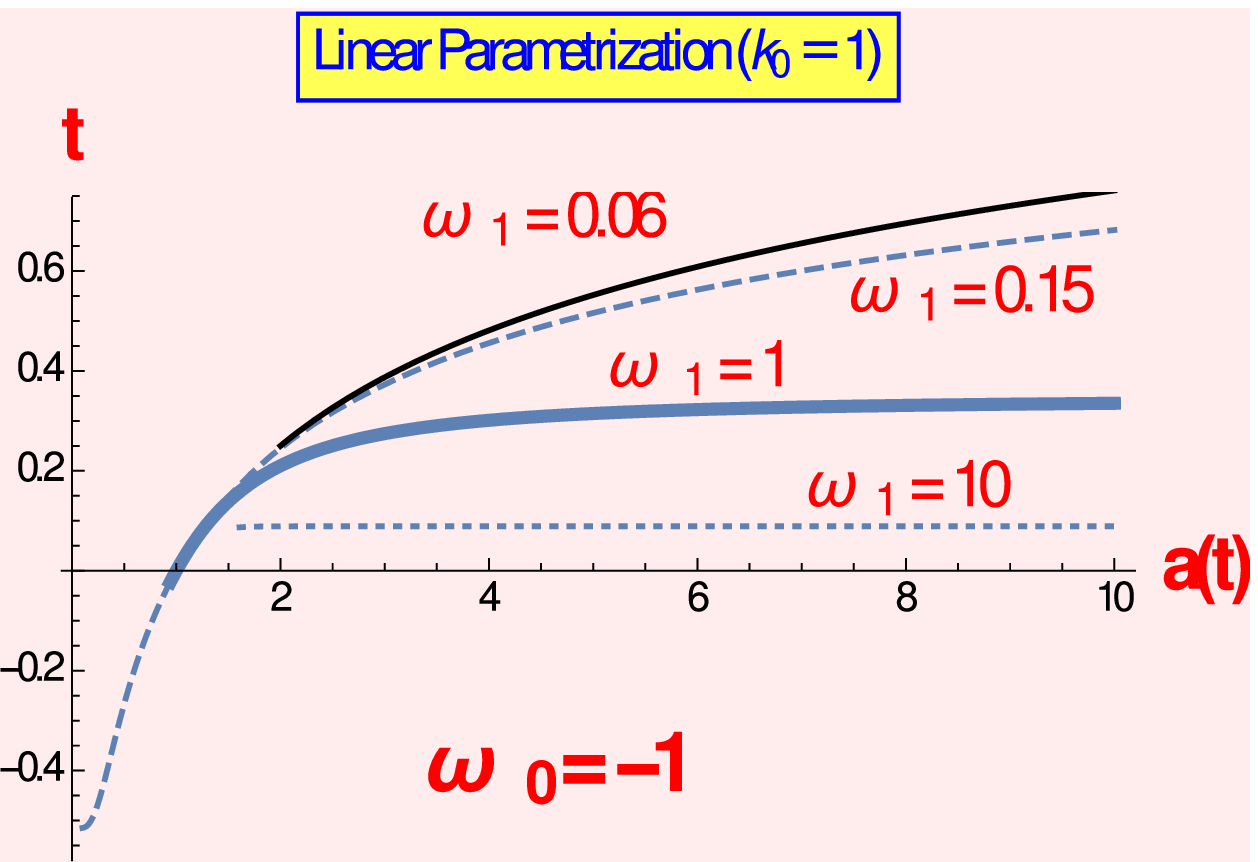}~~~~~\includegraphics[height=1.7in, width=2.2in]{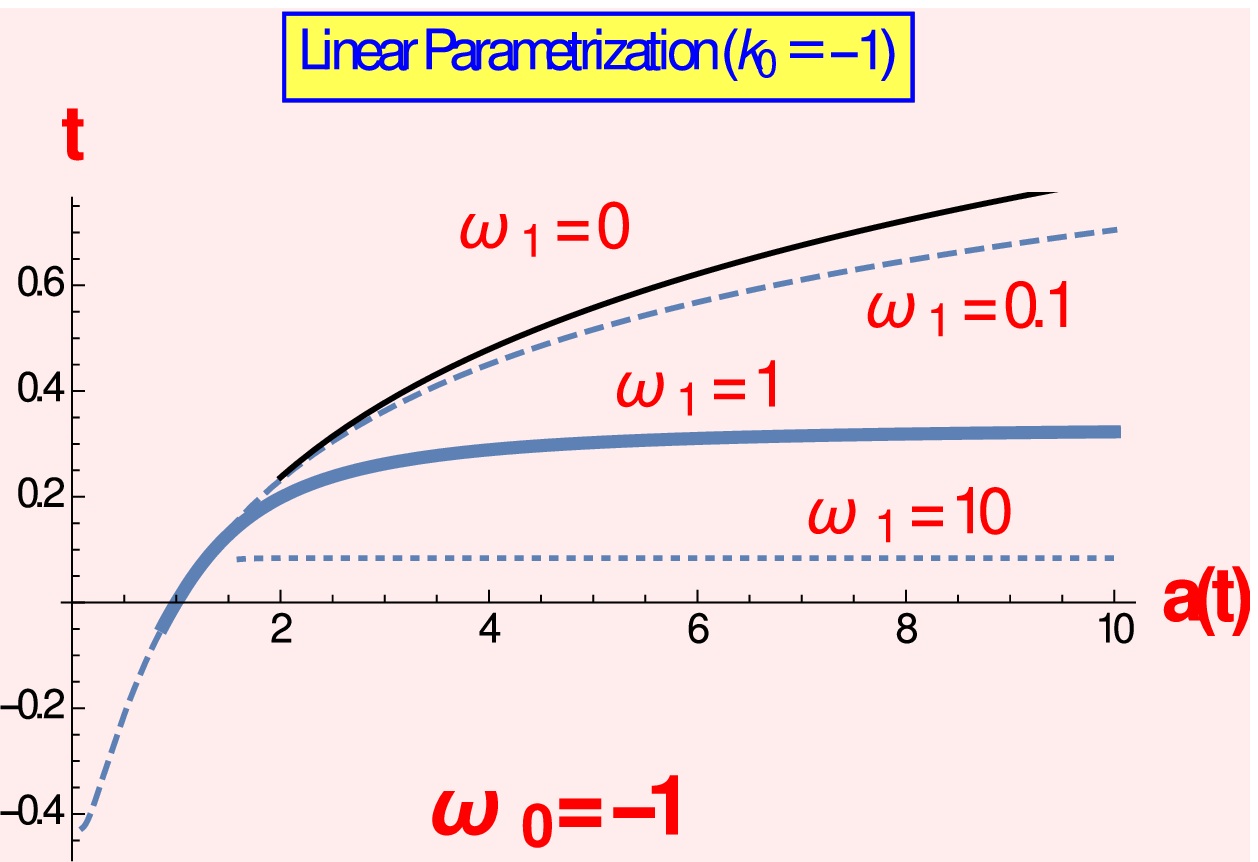}\\
Fig. 1. Plots of $t$ w.r.t. $a(t)$ for linear parametrization for $k_0 = 0$  and $1 , -1$ respectively. \\
\end{center}
\end{figure}
We see as time increases scale factor becomes steeply increasing. If we increase the value of $\omega_1$ keeping $\omega_0 = -1$ constant, for lower values of time, scale factor blows up quickly. Now, $z$ goes negative with respect to time. If time increases, $z$ decreases and $\omega_0 + \omega_1z$ also decreases.  This will create negative pressure which will be effected as the increment of the value of the scale factor.

Similarly, for \textbf{CPL redshift parametrization} we get from (\ref{CPL_redshift_parametrization}) and (\ref{Mass_Scale_Factor_Relation}) ,
\begin{equation}\label{CPL_II}
M = \rho^{\frac{1}{1+\omega_0+{\frac{\omega_1 z}{1+z}}}} = (m_0a^{-3}) \Rightarrow M^{1+\omega_0+{\frac{\omega_1 z}{1+z}}} = \rho = (m_0a^{-3})^{1+\omega_0+{\frac{\omega_1 z}{1+z}}}. \\
\end{equation}
So
\begin{equation}\label{CPl_III}
 p(z)= \left\{\omega_0+\frac{\omega_1 z}{1+z}\right\}(m_0a^{-3})^{1+\omega_0+{\frac{\omega_1 z}{1+z}}} 
\end{equation} 
Now from (\ref{field_equation_II}),
\begin{equation}\label{adot_for_CPL}
\left(\frac{dt}{da}\right) = \left[ 2\left\{\frac{4\pi}{3} (m_0a^{-3})^{1+\omega_0+{\frac{\omega_1 z}{1+z}}}  + \frac{\Lambda}{6}\right\} a^2 - k_0 \right]^{-\frac{1}{2}} 
\end{equation}
solving $t$ for $a(t)$, 
\begin{equation}\label{t-a_for_CPL}
t - t_0 = \int \left[ 2\left\{\frac{4\pi}{3} (m_0a^{-3})^{1+\omega_0+{\frac{\omega_1 z}{1+z}}}  + \frac{\Lambda}{6}\right\} a^2 - k_0 \right]^{-\frac{1}{2}} da 
\end{equation}
 
\begin{figure}[h!]
\begin{center}

(a)~~~~~~~~~~~~~~~~~~~~~~~~~~~~~~~~~~~~~~~~~~~~~(b)~~~~~~~~~~~~~~~~~~~~~~~~~~~~~~~~~~~~~~~~~~(c)
\includegraphics[height=1.7in, width=2.2in]{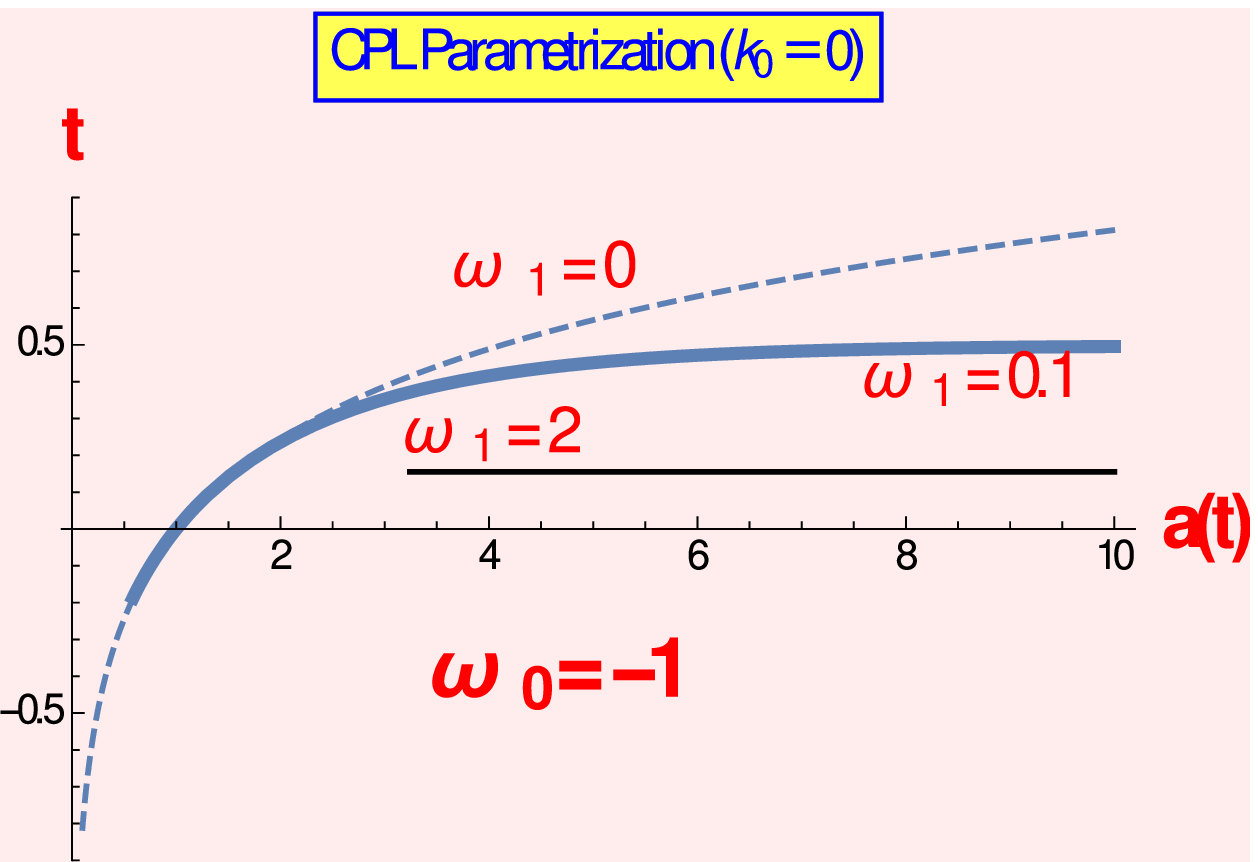}~~~~\includegraphics[height=1.7in, width=2.2in]{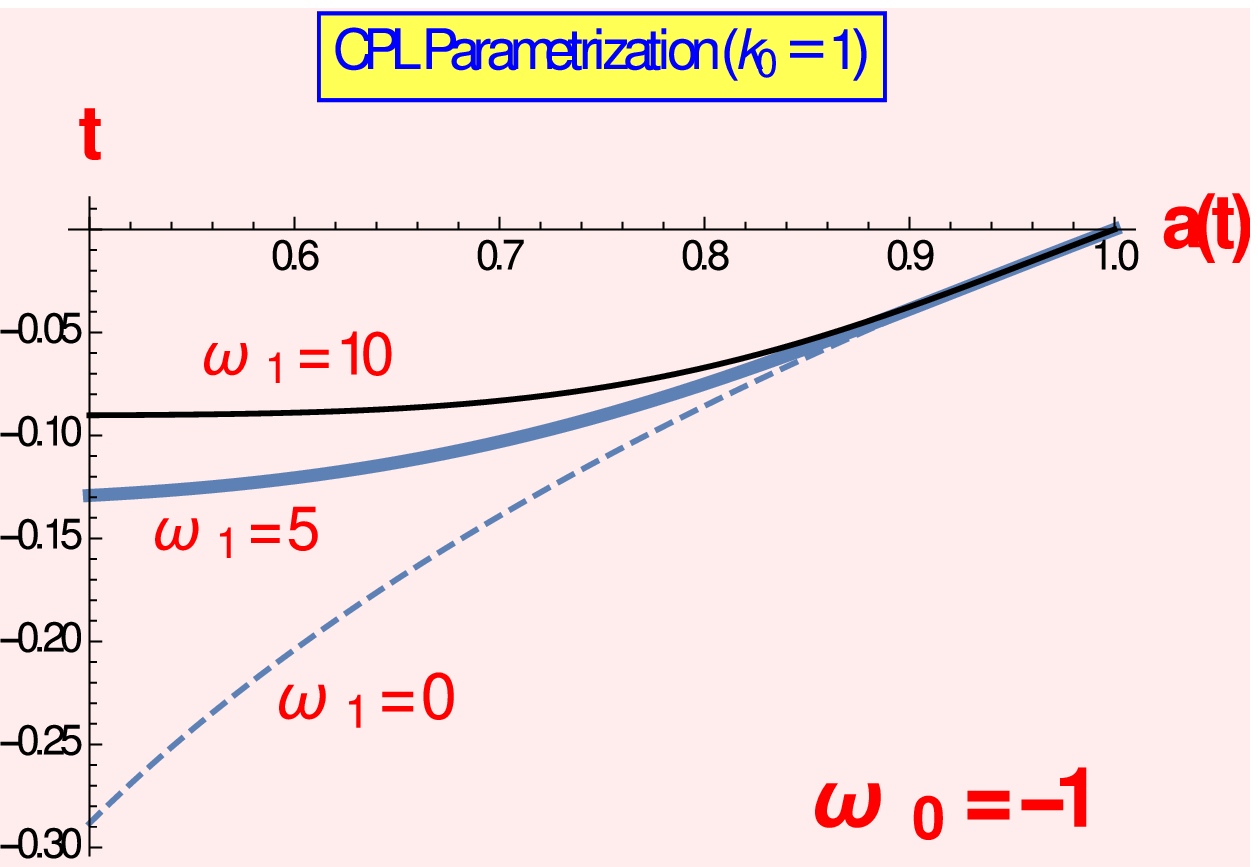}~~~~\includegraphics[height=1.7in, width=2.2in]{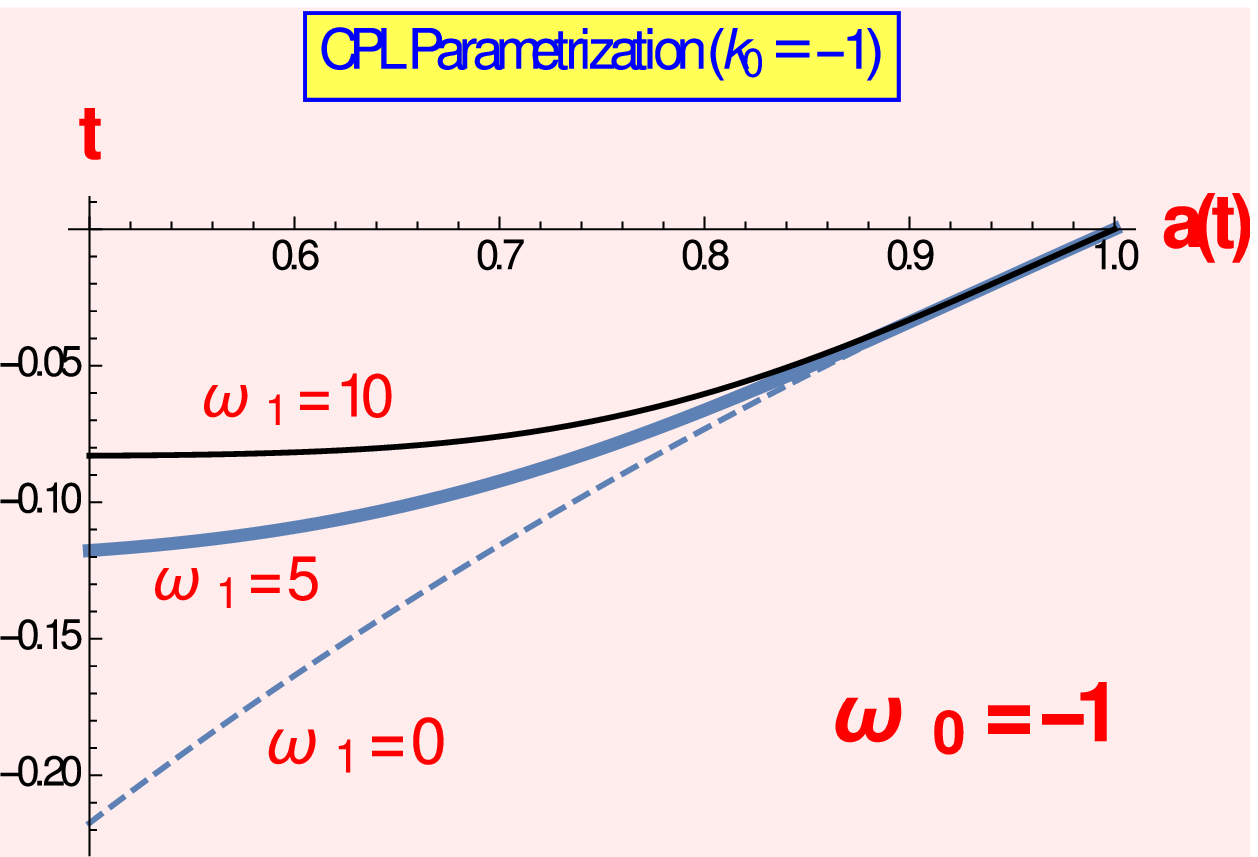}\\
Fig. 2. Plots of $t$ w.r.t. $a(t)$ for CPL parametrization for $k_0 = 0$  and $1 , -1$ respectively.
\end{center}
\end{figure}

Here we have plotted the changes of time with respect to the scale factor keeping $\omega_0=-1$ always. 

We plot the $t-a(t)$ graphs for $k_0= 0$ in Fig. 2(a). We plot $\omega_1= 0$, $0.1$ and $2$ cases. In all of the cases we see the increment of $t$ is high for small $a(t)$ and as $a(t)$ is increased, the rate of increment of $t$ slows down. The rate of this slowing down is again high for high $\omega_1$ values.

For Fig. 2(b), here we plot $\omega_1= 0$, $5$ and $10$ cases. We can see that increasing of the value of scale factor, the increment of time increases rapidly. For lesser values of the scale factor the time is noticibly different for $\omega_1$ values. But for greater values of the scale factor, we observe the times vs scale factor curves are almost same. Here this will also create negative pressure for less value of 1. For the range of $\omega_1$ is $-0.145 < \omega_1 < -0.12$ the increment of time goes strictly high in $t > -0.04 $ where the scale factor $a(t)> 0.9$.

For Fig. 2(c), here is just same phenomena happen, but the time region slightly changes. As we have choosen same values of $\omega_1$ as Fig. 2(b), we can see the time increment range is $-0.11 < t < -0.10$.

More simply we can state this incident as the distance between the $a(t)$ axis and $t$ vs $a(t)$ curve at a particular $a(t)$ is shorter if we choose higher $\omega_1$. To analyse this incident physically we will recall the fact that the redshift goes negative for a positive time. So increment in time makes $\omega_0 + \frac{\omega_1 z}{1+z} = ( \omega_0 + \omega_1 ) - \frac{\omega_1}{1+z} = (-1+\omega_1)- \frac{\omega_1}{1+z}$ to decrease with the increment of $\omega_1$ (keeping $\omega_0= -1$). So this causes more negative pressure and ultimately the concerned universe reaches to a future singularity more rapidly. We can conclude that the CPL model rapidly moves towards a future singularity than $`\Lambda$' model.

Similarly, for \textbf{JBP redshift parametrization} we get from (\ref{JBP_redshift_parametrization}) and (\ref{Mass_Scale_Factor_Relation}),
\begin{equation}\label{JBP_I}
M = \rho^{\frac{1}{\omega_0+\frac{\omega_1z}{(1+z)^2}+1}} = (m_0a^{-3}) \Rightarrow M^{\omega_0+\frac{\omega_1z}{(1+z)^2}+1} = \rho = (m_0a^{-3})^{{\omega_0+\frac{\omega_1z}{(1+z)^2}+1}}
\end{equation}
So
\begin{equation}\label{JBP_II}
p(z)= \left\{\omega_0+\frac{\omega_1z}{(1+z)^2}\right\}(m_0a^{-3})^{{\omega_0+\frac{\omega_1z}{(1+z)^2}+1}}
\end{equation}
Now from (\ref{field_equation_II}) 
\begin{equation}\label{adot_for_JBP}
\left(\frac{dt}{da}\right) = \left[ 2\left\{\frac{4\pi}{3} (m_0a^{-3})^{{\omega_0+\frac{\omega_1z}{(1+z)^2}+1}}  + \frac{\Lambda}{6}\right\} a^2 - k_0 \right]^{-\frac{1}{2}} 
\end{equation}
Solving $t$ for $a(t)$,
\begin{equation}\label{t-a_for_JBP}
\Rightarrow t - t_0 = \int \left[ 2\left\{\frac{4\pi}{3} (m_0a^{-3})^{{\omega_0+\frac{\omega_1z}{(1+z)^2}+1}}  + \frac{\Lambda}{6}\right\} a^2 - k_0 \right]^{-\frac{1}{2}} da
\end{equation}
\begin{figure}[h!]
\begin{center}

~~~(a)~~~~~~~~~~~~~~~~~~~~~~~~~~~~~~~~~~(b)~~~~~~~~~~~~~~~~~~~~~~~~~~~~~~~~~~(c)
\includegraphics[height=1.7in, width=2.2in]{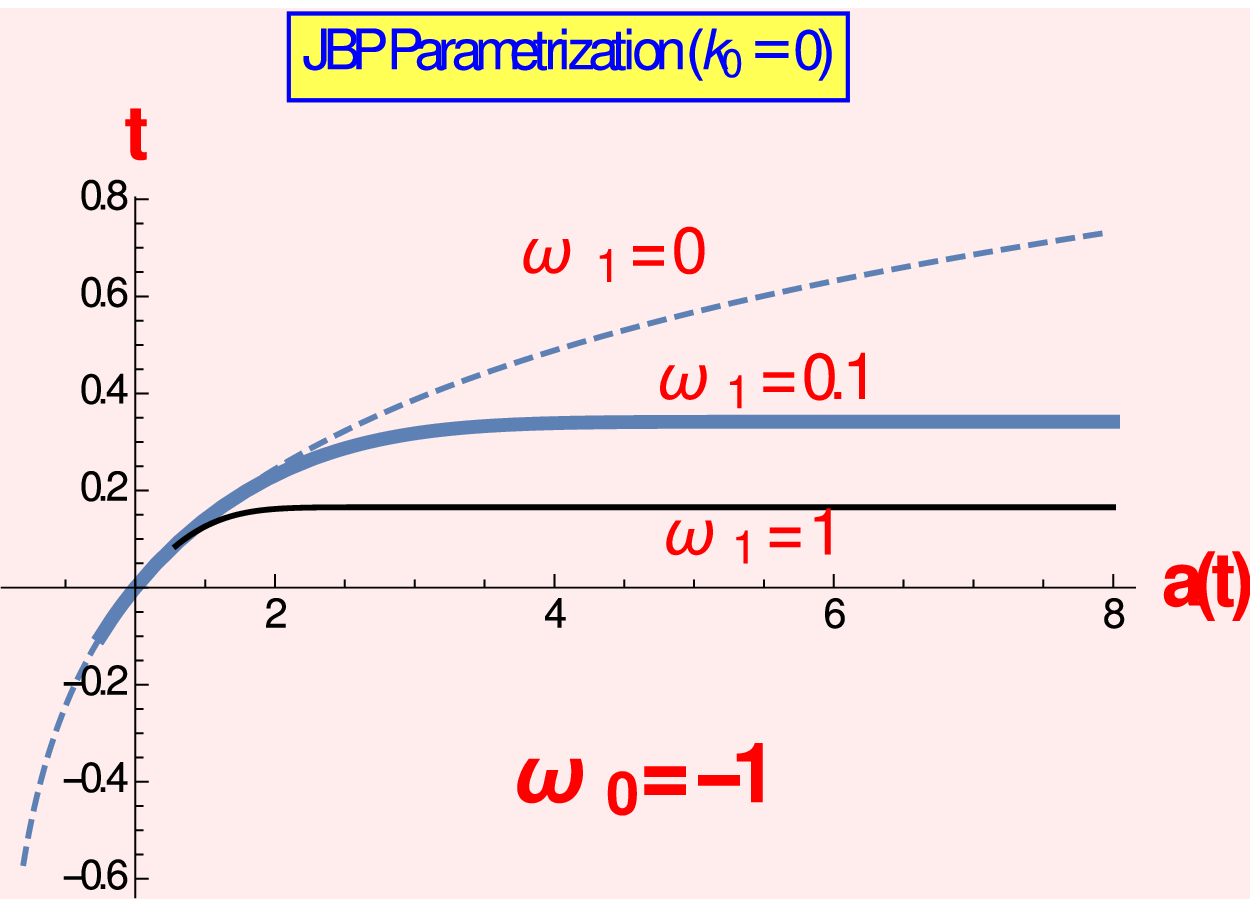}~~~~\includegraphics[height=1.7in, width=2.2in]{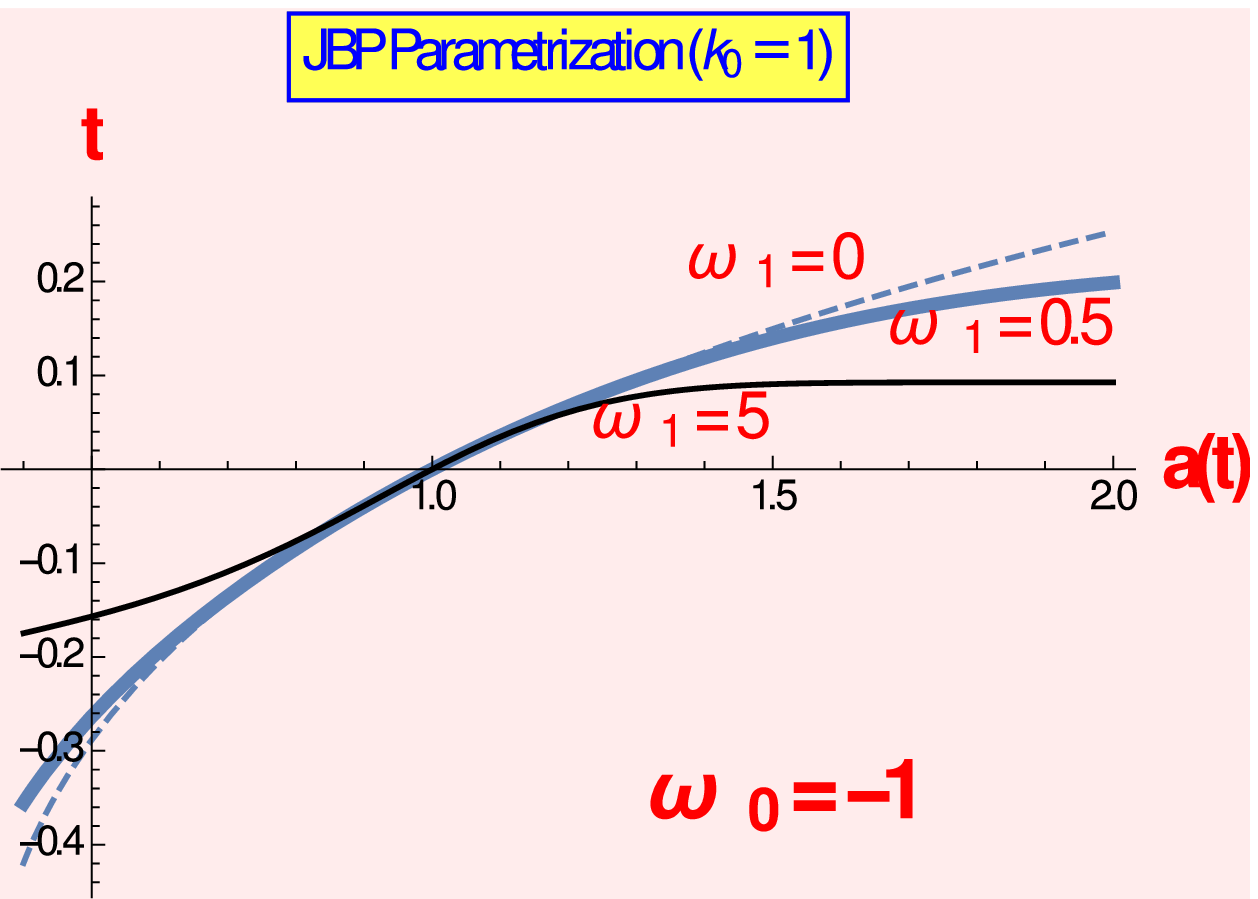}~~~~\includegraphics[height=1.7in, width=2.2in]{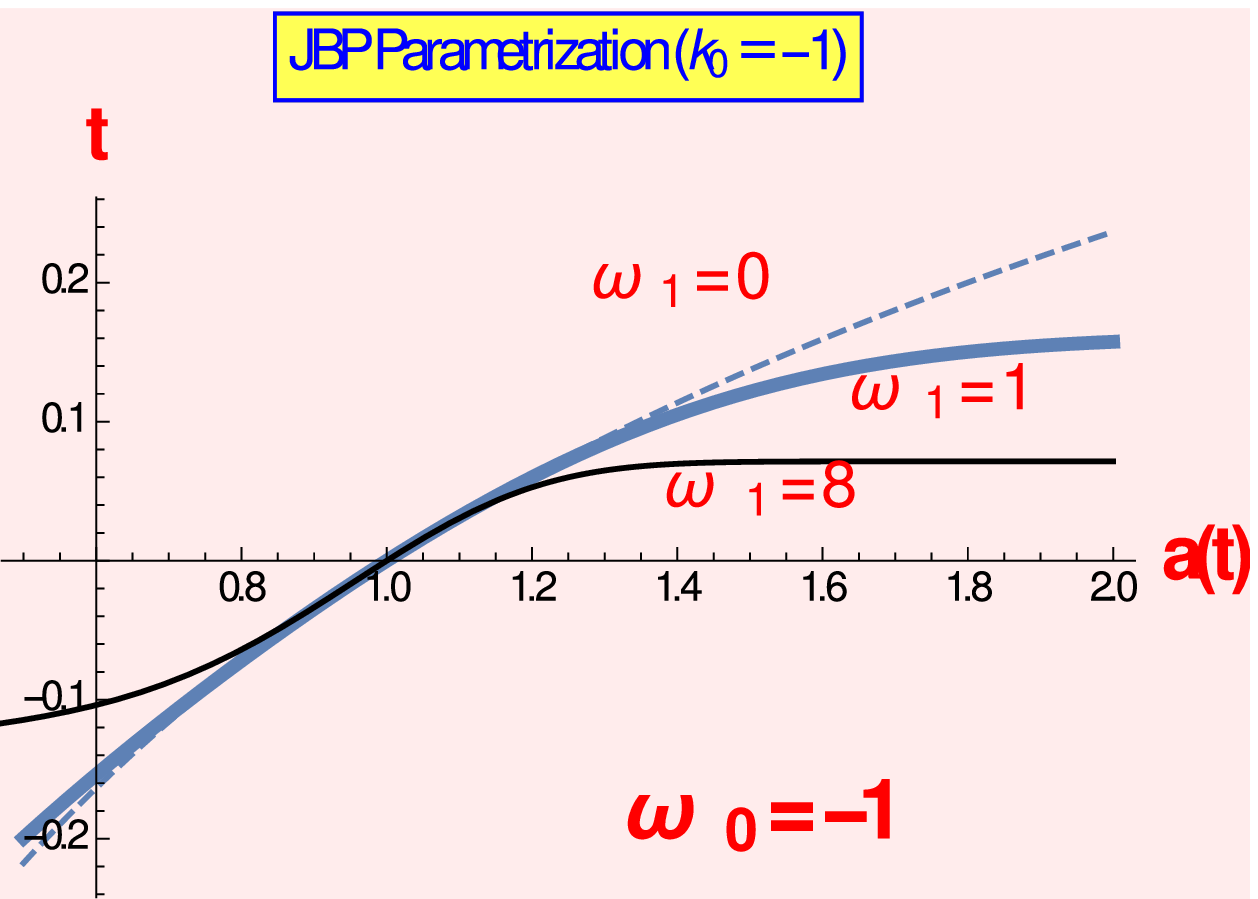}\\
Fig. 3. Plots of $t$ w.r.t. $a(t)$ for JPB parametrization for $k_0 = 0$  and $1 , -1$ respectively.
\end{center}
\end{figure}

Now for JBP redshift parametrization we will analyse three Figures for $k_0 = 0, 1, -1$ keeping $\omega_0$ constant.

In the Fig. 3(a), whrere $k_0= 0$ and $\omega_0 =-1$, We plot $\omega_1= 0$, $0.1$ and $1$ cases. The graph of time increment steeply decreases for ascending different values of $\omega_1$. For the range of $\omega_1$ $0.1< \omega_1 \leq 1$, the increment of time tends to remain constant.

In the Fig. 3(b), we plot $\omega_1 = 0$, $0.5$ and $5$ cases. For $k_0= 1$, at first, the graph of time increment increases then it remains constant in a region and finally decreases accordingly for the ascending values of $\omega_1$. In the time region $-0.08 < t < 0.05$, it remains constant for all values of $\omega_1$ when the scale factor region is $0.9 < a(t) < 1.2$.

In the Fig. 3(c), for $k_0= -1$, the difference of the time increment with respect to scale factor is nearly same as Fig. 3(b), but the time region where the increment remains constant is $-0.08 < t < 0.05$ for scale factor range $0.85 < a(t) < 1.05$.

Therefore, in JBP parametrization, we can see that the rate of time increment is slowly down and remains almost constant for higher values of $a(t)$.

Now, for \textbf{Logarithmic Redshift parametrization} we get from (\ref{Logarithmic_redshift_parametrization}) and (\ref{Mass_Scale_Factor_Relation}),

\begin{equation}\label{Logarithmic_redshift_I}
M^{\omega_0+\omega_1\ln(1+z)+1} = \rho = ({m_0a^{-3}})^{\omega_0+\omega_1\ln(1+z)+1} 
\end{equation}
Now from (\ref{field_equation_II})  
\begin{equation}\label{adot_for_Logarithmic_redshift}
\left(\frac{dt}{da}\right) = \left[ 2\left\{\frac{4\pi}{3} ({m_0a^{-3}})^{\omega_0+\omega_1\ln(1+z)+1}  + \frac{\Lambda}{6}\right\} a^2 - k_0 \right]^{- \frac{1}{2}}
\end{equation}
Solving $t$ for $a(t)$,
\begin{equation}\label{t-a_for_Logarithmic_redshift}
t - t_0 = \int \left[ 2\left\{\frac{4\pi}{3} ({m_0a^{-3}})^{\omega_0+\omega_1\ln(1+z)+1}  + \frac{\Lambda}{6}\right\} a^2 - k_0 \right]^{-\frac{1}{2}} da
\end{equation}
We now plot three Figures for $k_0= 0, 1, -1$ in 4a, 4b, 4c
\begin{figure}[h!]
\begin{center}
~~~~4a~~~~~~~~~~~~~~~~~~~~~~~~~~~~~~~~~~~4b~~~~~~~~~~~~~~~~~~~~~~~~~~~~~~~~~~~4c
\includegraphics[height=2in, width=2.2in]{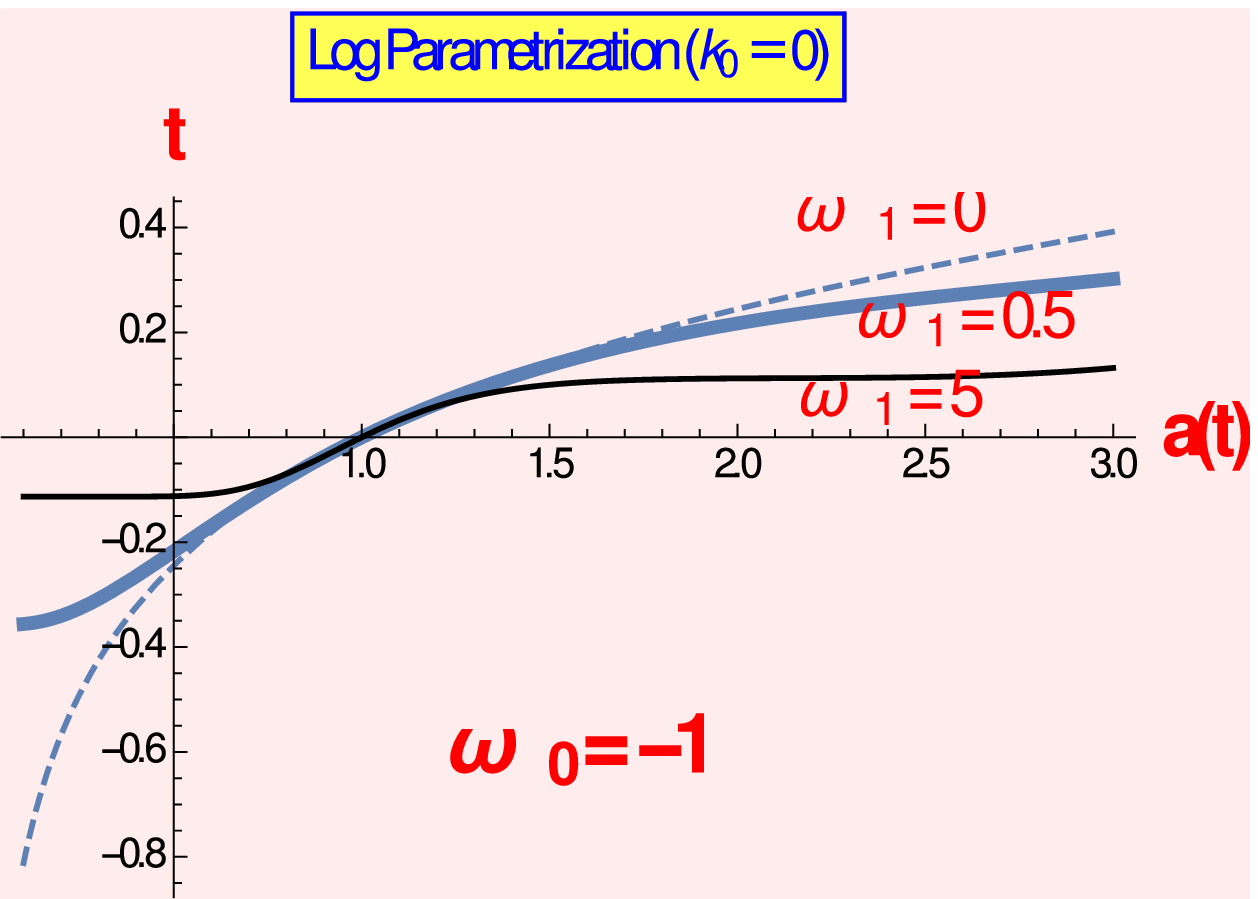}~~~~
\includegraphics[height=2in, width=2.2in]{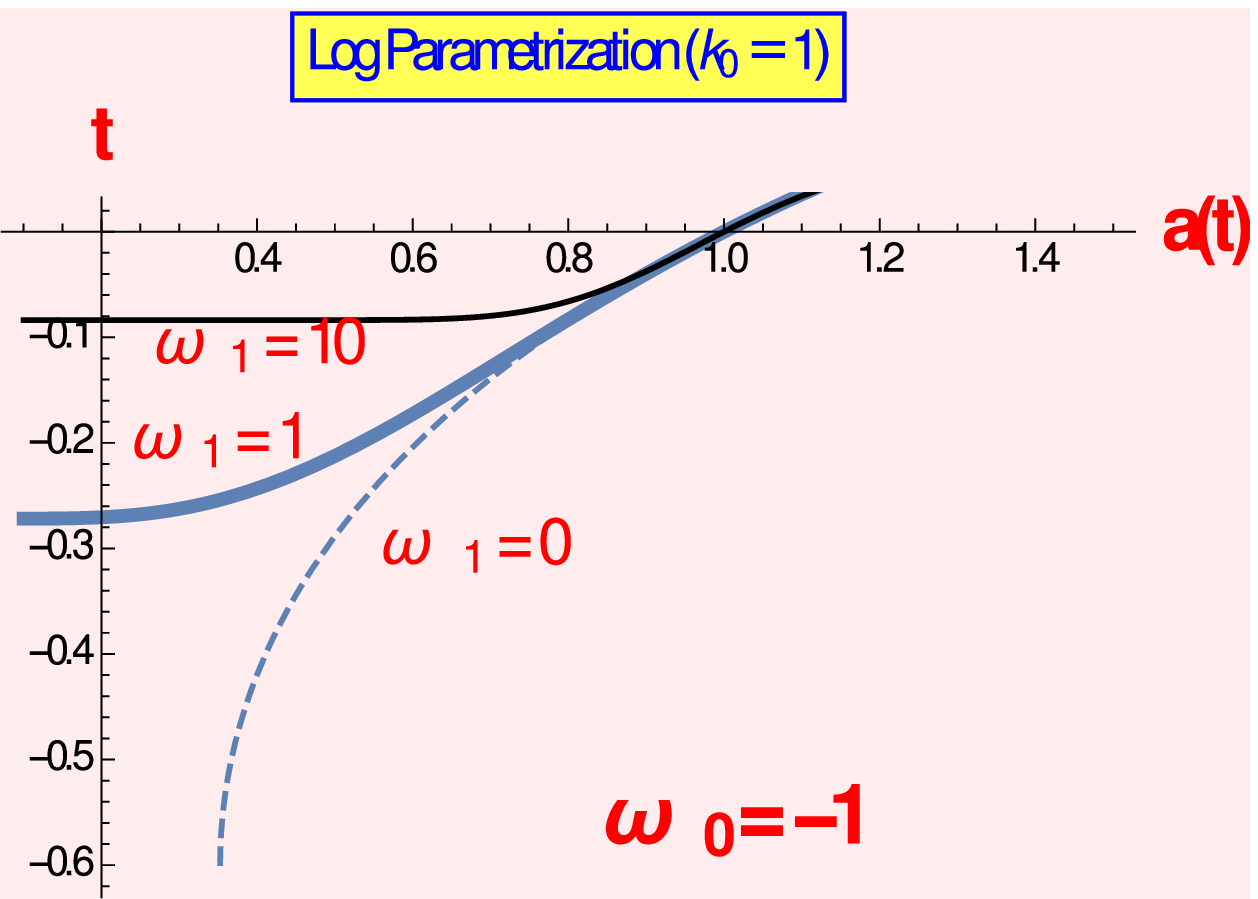}~~~~
\includegraphics*[height=2in, width=2.2in]{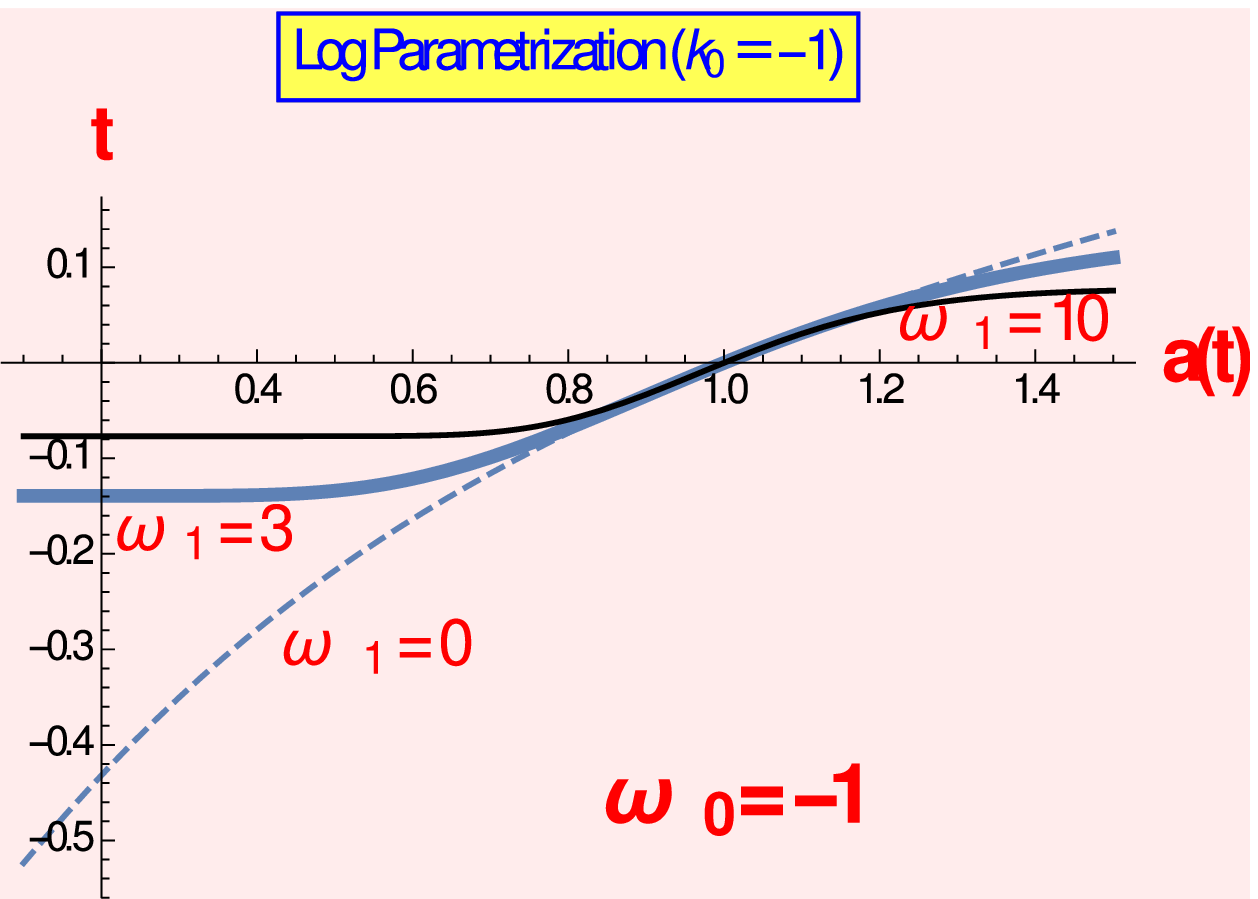}\\
Fig 4a-4c : Plots of $t$ w.r.t. $a(t)$ for Logarithmic redshift parametrization for $k_0 = 0$  and $1 , -1$ respectively.
\end{center}
\end{figure}

For Fig. 4(a), keeping $\omega_0= -1$, we can see the time increment graph starts from negative and then it remains same for all values of $\omega_1$ and after that it decreases in positive scale. Here we can also notice a region where the increment remains same. That time region range is $-0.08 < t < 0.05$ for the range of scale factor is $0.9 < a(t) < 1.2$. For $\omega_0= 0$, the $t-a(t)$ curve is noticibly different from others. At higher values of $t$, $a(t)$ diverges.

For Fig 4(b), keeping $\omega_0= -1$, The increment of time was different and it increases according as the higher values of $\omega_1$. But for $a(t) > 0.9$, the time increment is constant for all values of $\omega_1$. For high $t$ and high $\omega_1$, the $a(t)$ graph is increasing.  

For Fig. 4(c), where $k_0= -1$, the same phenomena like Fig. 4(a) happen. In the range $0.9 < t < 1.15$ the time increment is same for all values of $\omega_1$.

For \textbf{ASSS parametrization} we get from (\ref{ASSS_redshift_parametrization}) and (\ref{Mass_Scale_Factor_Relation}), 

\begin{equation}\label{ASSS_I}
M^{\frac{3}{(1+z)}\frac{A_0+2A_1(1+z)+A_2(1+z)^2}{A_1+2A_2(1+z)}} = \rho = ({m_0a^{-3}})^{\frac{3}{(1+z)}\frac{A_0+2A_1(1+z)+A_2(1+z)^2}{A_1+2A_2(1+z)}} 
\end{equation}
then from (\ref{field_equation_II}) 
\begin{equation}
\left(\frac{dt}{da}\right) = \left[ 2\left\{\frac{4\pi}{3} ({m_0a^{-3}})^{\frac{3}{(1+z)}\frac{A_0+2A_1(1+z)+A_2(1+z)^2}{A_1+2A_2(1+z)}}  + \frac{\Lambda}{6}\right\} a^2 - k_0 \right]^{-\frac{1}{2}} 
\end{equation}
Solving $t$ for $a(t)$,
\begin{equation}\label{t-a_for_ASSS}
t - t_0 = \int \left[ 2\left\{\frac{4\pi}{3} ({m_0a^{-3}})^{\frac{3}{(1+z)}\frac{A_0+2A_1(1+z)+A_2(1+z)^2}{A_1+2A_2(1+z)}}  + \frac{\Lambda}{6}\right\} a^2 - k_0 \right]^{-\frac{1}{2}} da
\end{equation}

\begin{figure}[h!]
\begin{quotation}
\begin{center}
(a)~~~~~~~~~~~~~~~~~~~~~~~~~~~~~~~~~~~~~~~~~~~~(b)~~~~~~~~~~~~~~~~~~~~~~~~~~~~~~~~~~~~~~~~~~~~~~~~~(c)
\includegraphics[height=1.7in, width=2.2in]{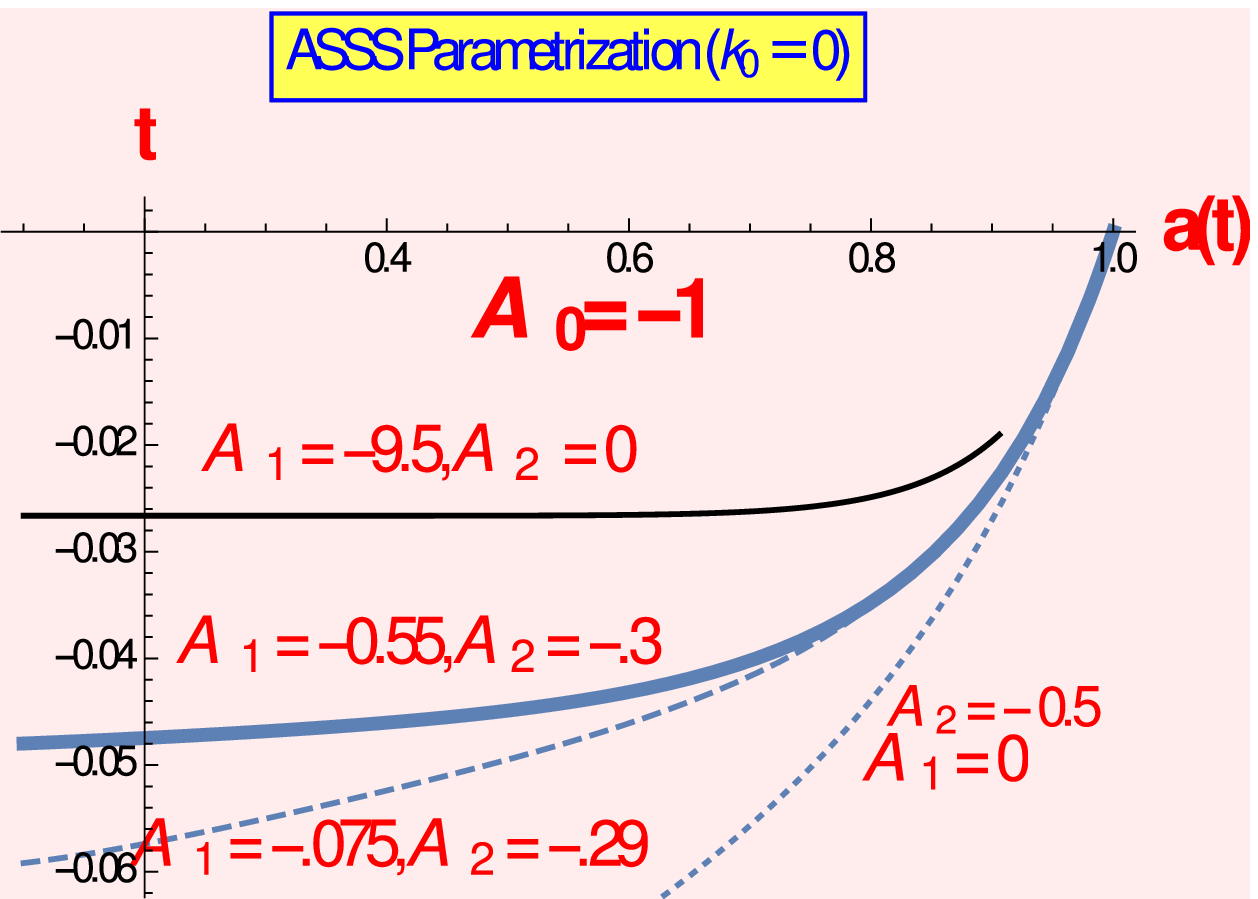}~~~~
\includegraphics[height=1.7in, width=2.2in]{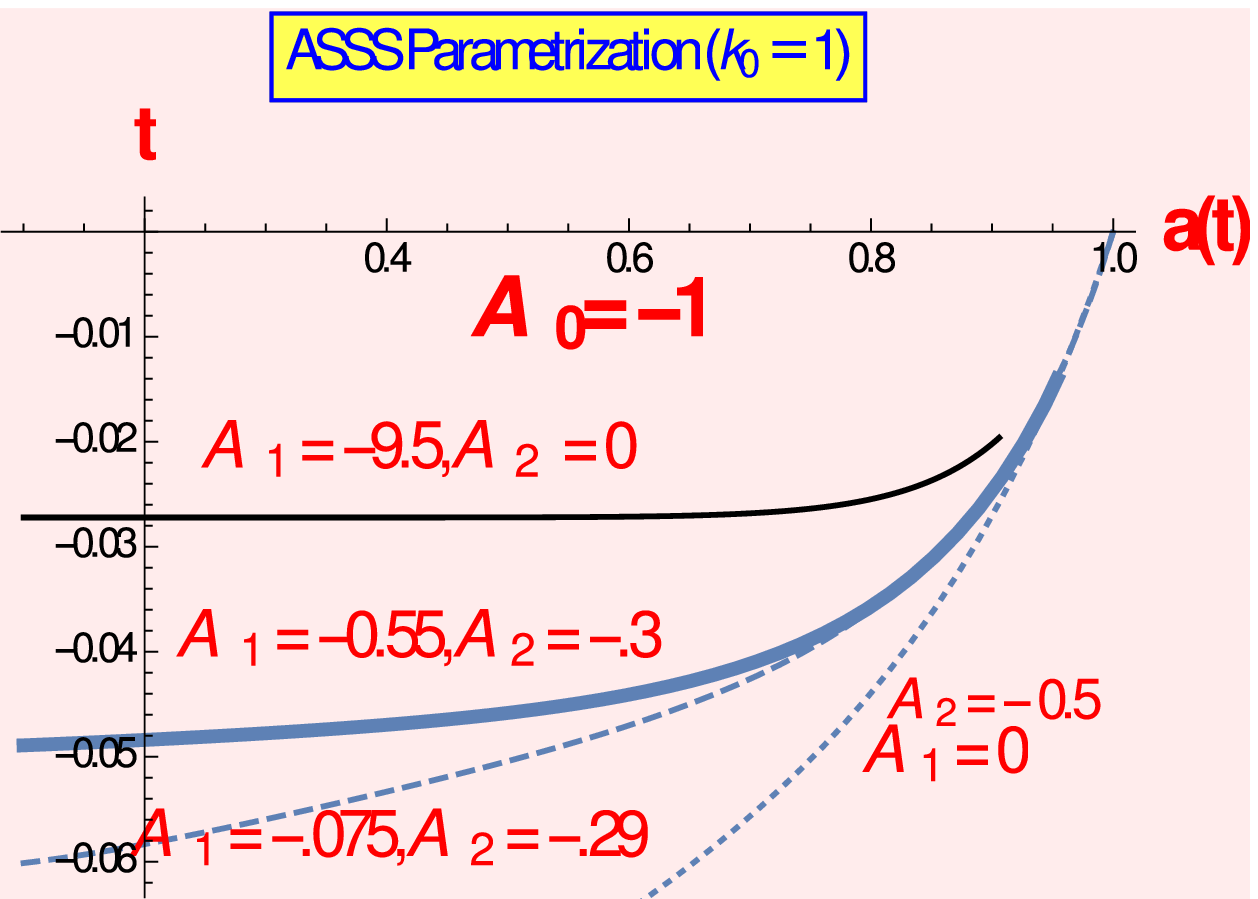}~~~~
\includegraphics[height=1.7in, width=2.2in]{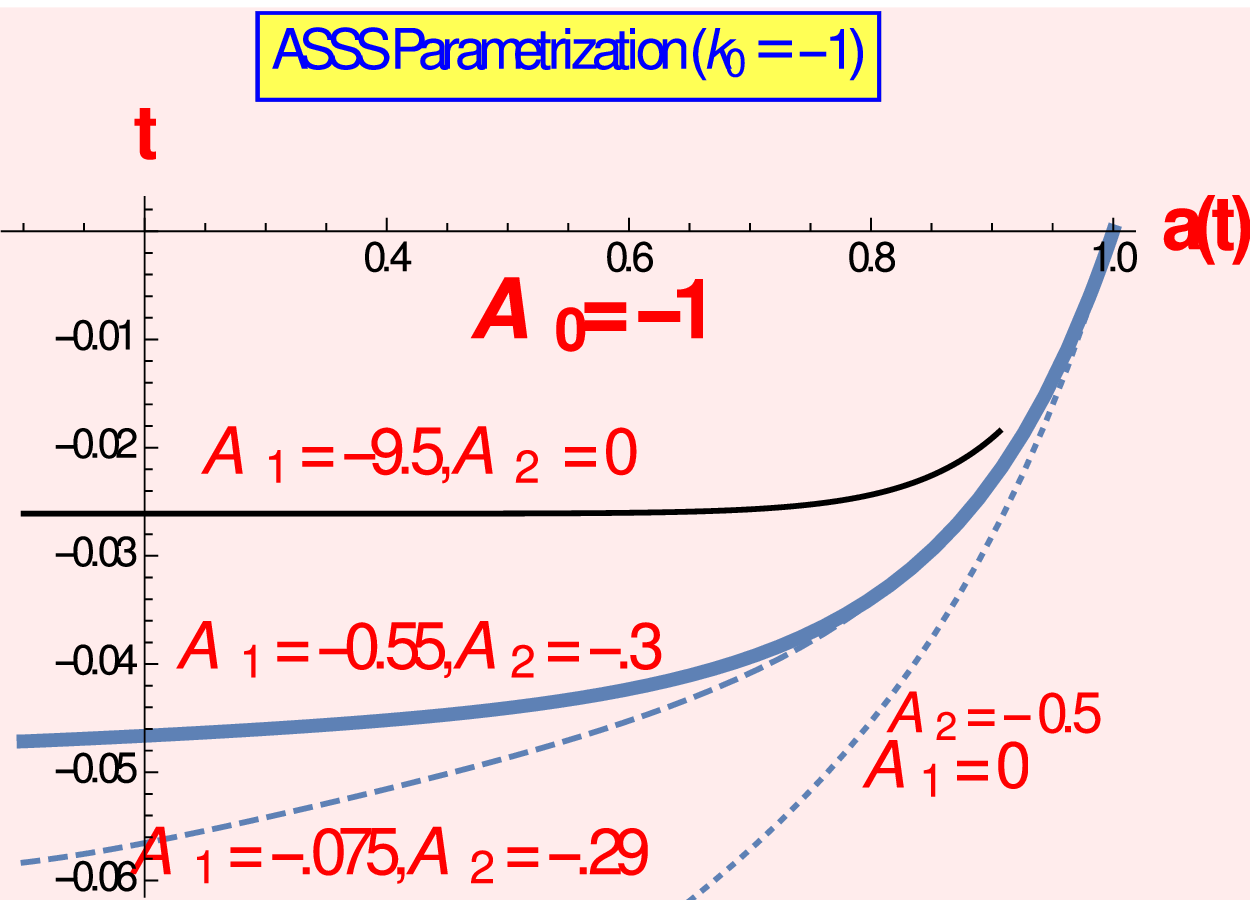}~~~~\\
Fig. 5.  Plots of $t$ w.r.t. $a(t)$ for ASSS parametrization for $k_0 = 0$  and $1 , -1$ respectively.
\end{center}
\end{quotation}
\end{figure}
In ASSS parametrization the cases for $k_0= 0, 1, -1$ are nearly same. The time increment is strictly high and constant, when the scale factor $a(t) > 0.95$. 

For ASSS, we plot $t$ vs $a(t)$ in the Fig. 5(a) to 5(c). For the cases of low $|A_1|$ and $|A_2|$, i.e., which may look similar to the $\Lambda$CDMcase, we see the slope of $t$ vs $a(t)$ is high. But as we increase the values of $|A_1|$ and $|A_2|$, for low $a(t)$ weget a $t$ curve which is almost parallel to the $a(t)$ axis. More we increase the values of $|A_1|$ and $|A_2|$, more parallel is the $t-a(t)$ curve for low $a(t)$. But in later phase (i.e., for high $a(t)$), we see the graphs to erge with the low $|A_1|$ and $|A_2|$ cases. Physically this incident can be interpreted as: for low $t$, i.e., for the early phase of universe has grown older, $a(t)$ converges to a finite value. This model does not support the big rip theory. However, for $k_0= 1$( Fig 5(b)) and $k_0= -1$ (Fig. 5(c)), we get almost the same nature as the $k_0= 0 $ case.

Similarly, for \textbf{GCCG models} ,

from (\ref{GCCG_EoS}) and (\ref{field_equation_II}) we have
\begin{equation}\label{adot_for_GCCG}
\left(\frac{da}{dt}\right) = \left[ 2{a^2}\left\{\frac{4\pi}{3} \left[\left\{({m_0a^{-3}})^{(\omega+1)(\alpha+1)}-1\right\}^{\frac{1}{\omega + 1}} + C \right]^{\frac{1}{\alpha + 1}} + \frac{\Lambda}{6}\right\}  - k_0 \right]^{\frac{1}{2}}
\end{equation}
Solving $t$ for $a(t)$,
\begin{equation}\label{t-a_for_GCCG} 
\Rightarrow t -t_0 = \int\left[ 2{a^2}\left\{\frac{4\pi}{3} \left[\left\{({m_0a^{-3}})^{(\omega+1)(\alpha+1)}-1\right\}^{\frac{1}{\omega + 1}} + C \right]^{\frac{1}{\alpha + 1}} + \frac{\Lambda}{6}\right\}  - k_0 \right]^{-\frac{1}{2}}
\end{equation}
We have calculated analytic function for rest of parametrizations, i.e., for \textbf{CPL}, \textbf{JBP} and \textbf{Log Parametrization} . We have found same results as \textbf{linear parametrization} for $k_0= 0, 1, -1$ and for \textbf{ASSS} and \textbf{GCCG}, we can not find any proper analytic value.
\begin{figure}[h!]
\begin{quotation}
\begin{center}
(a)~~~~~~~~~~~~~~~~~~~~~~~~~~~~~~~~~~~~~~~~~(b)~~~~~~~~~~~~~~~~~~~~~~~~~~~~~~~~~~~~~~~~~(c)
\includegraphics[height=1.7in, width=2.2in]{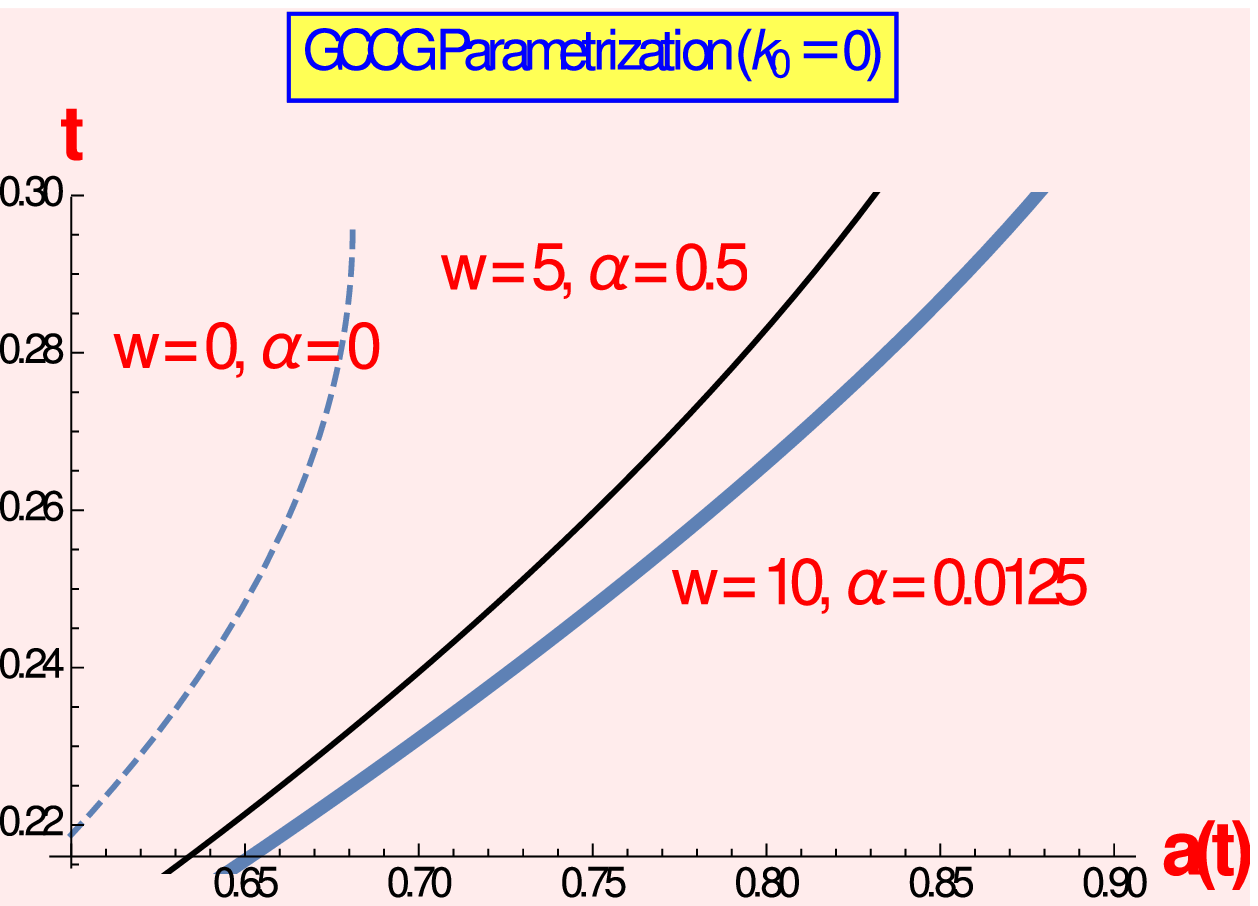}~~~~
\includegraphics[height=1.7in, width=2.2in]{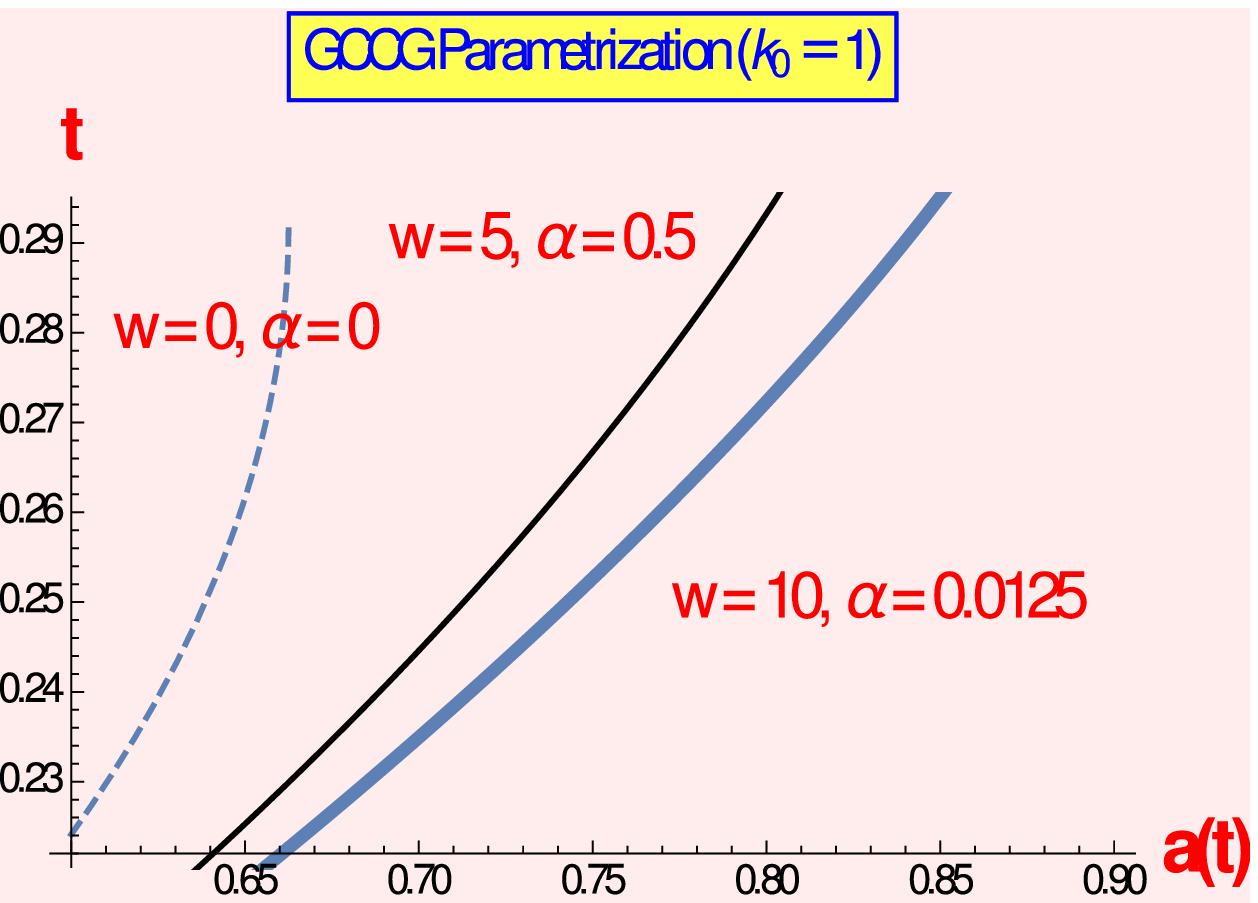}~~~~
\includegraphics[height=1.7in, width=2.2in]{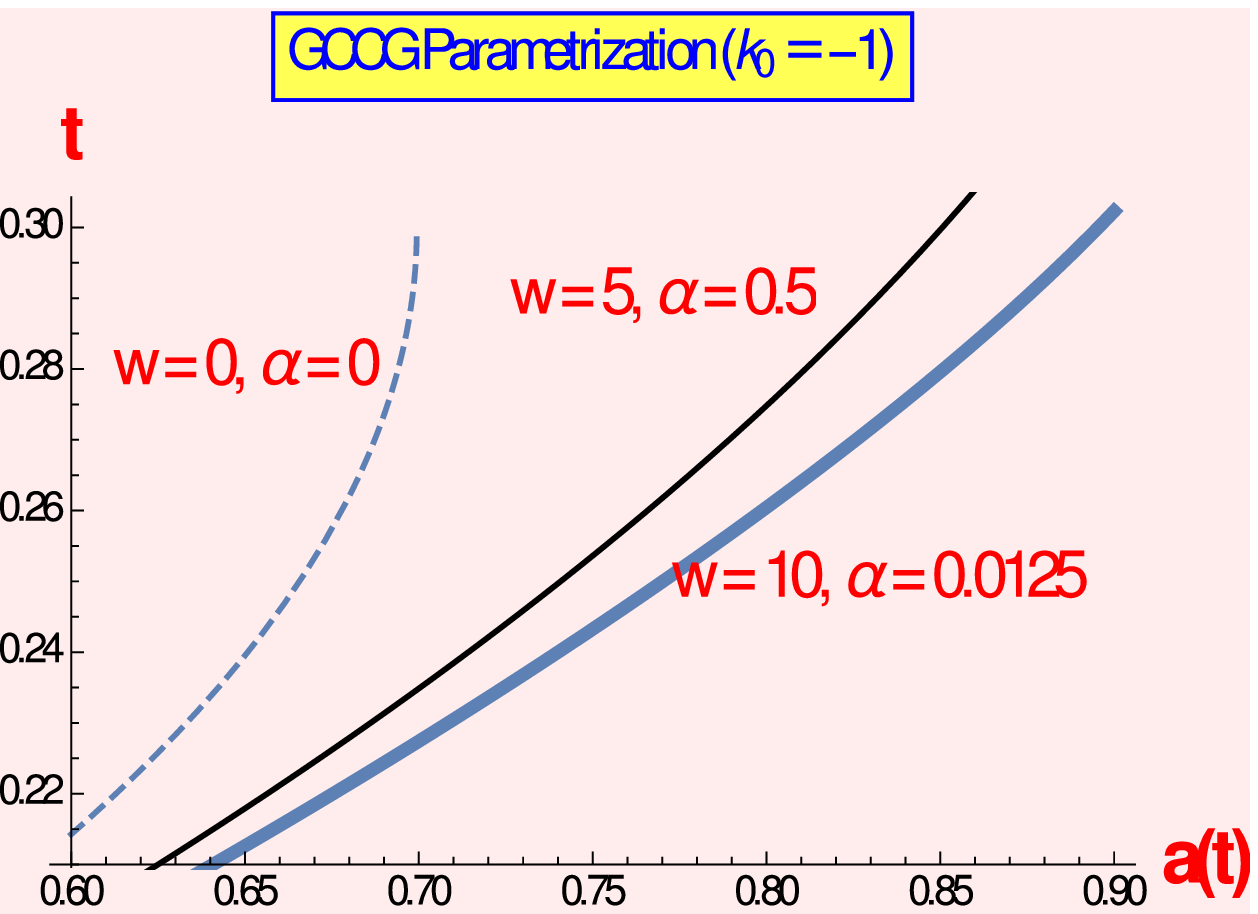}~~~~\\
Fig. 6.  Plots of $t$ w.r.t. $a(t)$ for GCCG parametrization for $k_0 = 0$  and $1 , -1$ respectively.
\end{center}
\end{quotation}
\end{figure}

For GCCG model, we can see the three cases are mostly same else their time range for the values of $\omega$ and $\alpha$.

Fig. 6(a) is a graph of $t$ vs $a(t)$ for flat universe. For $\Lambda$CDM case, i.e., when $\omega$=$\alpha$=$0$, we see time is an increasing function of $a(t)$. For low $a(t)$, $t$ grows slowly and later it increases rapidly and diverges to infinite $t$ at a finite $a(t)$. This means, for terminal GCCG case scale factor never blows to the infinity and always stay finite. As we increase $\omega$ and $\alpha$ we see the slope of $t-a(t)$ curve decreases. But we will have finite $a(t)$ for all the $t$ domain. The same trend is carried for $k_0=1$ (Fig. 6(b)) and $-1$ (Fig 6(c)) cases.We know GCCG does not allow a future singularity \cite{Pedro F_González_Díaz}. The graphs also do support the preestablished result.
\section{Brief Discussions and Conclusion}

Throughout this paper, we have studied the evolution of time t and scale factor $a(t)$ with respect to each other when the present day universe is supposed to experience a late-time cosmic acceleration. To interpret the effect of the negative pressure exerted by the exotic fluids filled in all over the universe homogeneously, we have taken different EoSs of such fluids which are dependent on the redshift z and some
arbitrary redshift parametrization parameters. Comparison between $\Lambda$CDM and $\omega(z)$CDM has been done in earlier works \cite{Nature_Astronomy}. We have shown in every of the models we have choosen that $\omega_0 = -1$ case ,i.e., the $\Lambda$-CDM cause allows $a(t)$ to grow for a large domain of t. This physically implicates that for $\Lambda$CDM we see an increasing scale factor with time but the rate of increment is not as high as these of the $\omega(z)$ parametrizations. Some other papers \cite{the_concordance_model} again do not consider a rigid $\Lambda$ term as a fundamental building block of the concordance $\Lambda$CDM model, where the energy density of universe is parametrized as a function of Hubble's rate. Firstly, we have discussed about the simplest model, namely, linear redshift parametrization. We see for all the values of $k_0 = -1, 0, 1$, if we take smaller $\omega_0$ and $\omega_1$ ,i.e., if we choose $\omega(z) = \omega_0 + \omega_1z$ to be of smaller value, the slope of the $t-a(t)$ curve is higher than the larger valued $\omega(z)$ cases (i.e., when $\omega_0$ and $\omega_1$ have higher value). This shows that linear parametrization parameter mainly supports strong cosmic accleration. However this may lead to a future singularity known as `` Big Rip ''. For CPL parametrization the tendency of $t$ vs $a(t)$ curve is same as that of the linear parametrization case. For CPL the slope of $a(t)$ vs $t$ is less steeper than the cases of linear parametrization, i.e., if we increase $t$ rate of $a(t)$ is not high as that increase $t$ the rate of increment of $a(t)$ is not high as that of the Linear Parametrization case. This signifies that for flat universe through CPL behaves like highly negative pressure exerting gas, the power of the negative pressure is less as comparable to the linear parametrization case. By another language we can say that for CPL, the theoritical negative pressure is more controllable than linear parametrization. For open or closed universe, this scenario changes completely. We see the $t$ vs $a(t)$ curves are almost constant for low $a(t)$, whereas they increase with the increment of $a(t)$. This says with $t$ increment of $a(t)$ takes place but under a finite range. For low $t$, high $a(t)$ is observed if $\omega_1$ is lesser. For those two later cases Big Rip does not occur in future . For JBP the curves are finite similar in shape with that of linear parametrization. But for a past time again we see a steep increment in $a(t)$ vs $\omega_1$ is increased. So we can speculate that JBP is good to describe a future as well as a past singularity together.For logarithmic parametrization the same trend of JBP carries on . For ASSS parametrization we see the proper curves are availablefor past time only. For GCCG, all these phenomena do not occure. As we increase $a(t)$ we see $t$ to blow up for low $\alpha$, low $\omega$. This says for even infinite time also we get a finite scale factor. As we increase $\omega$ we observe the curves to increase but the slope never becomes parallel to the $a(t)$ axis. This says at high $t$, $a(t)$ must converge to some finite values. This means no future singularity is allowed to be ocurred in GCCG model. This supports the purpose of proposition of such a fluid in \cite{Pedro F_González_Díaz}.  

\vspace{.1 in}
{\bf Acknowledgment:}
R.B. thanks IUCAA, Pune for Visiting Associateship. Both the authors thank Prof. Subenoy Chakraborty, Department of Mathematics, Jadavpur University for detailed discussions and thorough reviews of this paper.
  
\end{document}